# Crystal Growth and Magneto-transport of $Bi_2Se_3$ Single crystals

Geet Awana[1], Rabia Sultana[2], P. K. Maheshwari[2], Reena Goyal[2], Bhasker Gahtori[2], Anurag Gupta[2], and V.P.S. Awana[2,#]

[1]Department of Physics and Astrophysics Delhi University, New Delhi-110007, India
[2]CSIR-National Physical Laboratory, New Delhi-110012, India

In this letter, we report growth and characterization of bulk $Bi_2Se_3$ single crystals. The studied $Bi_2Se_3$ crystals are grown by self flux method through solid state reaction from high temperature (950˚C) melt of constituent elements and slow cooling (2˚C/hour). The resultant crystals are shiny and grown in [00l] direction, as evidenced from surface XRD. Detailed Reitveld analysis of PXRD (powder x-ray diffraction) of the crystals showed that these are crystallized in rhombohedral crystal structure with space group of R3m (D5) and the lattice parameters are a = 4.14(2) Å, b = 4.14 (2) Å and c = 28.7010(7) Å. Temperature versus resistivity ($\rho$-T) plots revealed metallic conduction down to 2K, with typical room temperature resistivity ($\rho_{300K}$) of around 0.53 m$\Omega$-cm and residual resistivity ($\rho_{0K}$) of 0.12 m$\Omega$-cm. Resistivity under magnetic field [$\rho$(T)H] measurements exhibited large +Ve magneto resistance right from 2K to 200K. Isothermal magneto resistance [$\rho$H] measurements at 2K, 100K and 200K exhibited magneto resistance (MR) of up to 240, 130 and 60% respectively at 14 Tesla. Further the MR plots are non saturating and linear with field at all temperature. At 2K the MR plots showed clear quantum oscillations at above say 10 Tesla applied field. Also the Kohler plots i.e., ($\Delta\rho/\rho_o$ versus B/ρ) were seen consolidating on one plot. Interestingly, the studied $Bi_2Se_3$ single crystal exhibited the Shubnikov-de Haas oscillations (SdH) at 2K under different applied magnetic fields ranging from 4Tesla to 14 Tesla

**[*]Corresponding Author**
Dr. V. P. S. Awana: E-mail: awana@nplindia.org
Ph. +91-11-45609357, Fax-+91-11-45609310
Homepage: awanavps.webs.com

## Introduction

Topological Insulators (TI) are a kind of wonder materials of topical interest today, where by the interior of the material is band insulating and the surface states are conducting [1,2]. Interestingly, the conducting surface states of the topological insulators are symmetry protected [3-5]. Further, the surface state carriers are quantized as their spins are locked in right angle to their momentum [1-5]. This gives rise to time reversal symmetry driven protected states. Clearly, the role of both spin and momentum of the protected surface states is important and hence the topological insulators are often called the futuristic potential spintronic materials [4-7]. No wonder, topological insulators with their rich physics and potential applications are the hottest topic today for condensed matter physicists including both theoreticians and experimentalists alike [1-7].

For experimental condensed matter physicists the work line starts from material, measurement to mechanism i.e., MMM. Basically, first and foremost task is to identify the master material and synthesize the same in right structure with best possible purity. The state of art measurements for various physical properties is the next step. Once the quality material is in place and the physical property characterization is over, one sits back and tries to analyze the results and explore the theoretical explanation and mechanism for the obtained physical property.

Keeping in view, the fact the topological insulators are the real hot cakes presently for condensed matter physicists and often in the beginning of a new field, the material quality is not always optimized to the best levels, we focus on growth and physical property characterization of by now one of the popular TI i.e., $Bi_2Se_3$. We did grow large (cm size) bulk $Bi_2Se_3$, exhibiting metallic character down to 2K and large +ve linear magneto resistance of up to 250% at 2K under applied field of 14 Tesla coupled with clear quantum oscillations above say 10 Tesla.

## Experimental Details

The $Bi_2Se_3$ sample is grown via self flux method [8] in an evacuated sealed quartz tube heating up to a temperature of 950 ˚C (2˚C/min) and then cooling down slowly to 650 ˚C (2˚C/hour). Both Bismuth (Bi) and Selenium (Se) powder were accurately weighed according to the stoichiometric ratio of 2:3 and then well mixed with the help of mortar and pestle. The above steps were performed in presence of high purity Argon atmosphere within a glove box (MBRAUN Lab star).The obtained mixed powder was then pelletized into a rectangular pellet form under a pressure of 50 kg/cm$^2$ by means of hydraulic press. After the pelletization was over the pellet was then sealed into an evacuated ($10^{-3}$Torr) quartz tube and placed inside the tube furnace. The furnace was heated up to 950˚C (2˚C/min) and then cooled very slowly to 650˚C (2˚C/hour), after which the same was switched off and allowed to cool naturally to room temperature. The structural characterization was performed through room temperature X-ray diffraction (XRD) using Cu-Kα radiation (λ=1.5418 Å). However, the magnetic measurements were done using the quantum design Physical Property Measurement System (PPMS). Also, the Scanning Electron Microscopy (SEM) and Energy Dispersive X-Ray Spectroscopy (EDAX) were carried out using ZEISS-EVO MA-10.

**Results and Discussions**

Figure 1 (a) and (b) represent the X-ray diffraction patterns of the synthesized single crystal $Bi_2Se_3$ sample. The on surface XRD pattern of single crystal $Bi_2Se_3$ clearly shows the 00l alignment, see Fig. 1(a). The rietveld refinement was carried out on powdered crystal using the FullProf suite Toolbar and the results are shown in Fig. 1(b). The synthesized $Bi_2Se_3$ sample exhibits a rhombohedral crystal structure with R3m (D5) space group. The lattice parameters as obtained from the Rietveld refinement are a = 4.14(2) Å, b = 4.14 (2) Å and c = 28.7010(7) Å and the values of α, β and γ are 90˚, 90˚ and 120˚.

Figure 2 represents the SEM image of the $Bi_2Se_3$ single crystal. The morphology of the synthesized $Bi_2Se_3$ single crystal exhibits layered structure, similar to that as reported by some of us recently for $Bi_2Te_3$ single crystals [8]. The left inset of Fig. 2 shows the compositional constituents of the studied $Bi_2Se_3$ single crystal. Only Bi and Se are seen, thus confirming the purity of the synthesized crystal. Further, this implies that the crystal is not contaminated by any abundant impurities like Carbon or Oxygen. The right inset of Fig. 2 depicts the quantitative weight% values of the atomic constituents (Bi and Se). The determined quantitative amounts of Bi and Se in studied $Bi_2Se_3$ crystal were found to be very near to stoichiometeric, i.e., close to $Bi_2Se_3$.

Figure 3 represents the percentage change of magneto-resistance (MR) under applied magnetic fields of up to 14 Tesla at various temperatures ranging from 2 K to 200 K. The MR is obtained using the equation MR = [R(H) - R(0)] / R(0). Figure 3 clearly shows that the synthesized $Bi_2Se_3$ crystal exhibits linearly increasing MR(%) values, reaching up to 240% at 2 K under applied magnetic field of about 14 Tesla. Also, oscillations are seen in MR above say 10 Tesla field at 2 K, which are marked by the arrows. These quantum oscillations can be referred to as the Shubnikov-de Haas (SdH) oscillations. More work is underway to quantify the observed MR oscillations. Clearly, the MR is linear and positive with quantum oscillatory changes at higher magnetic fields (> ~ 10 Tesla). The MR oscillatory behavior is observed at 2 K only, and at higher temperatures of 100 K and 200 K, though the MR is yet large enough and linear but the oscillations are not seen. The linear MR and low temperature oscillatory changes are in agreement with some of earlier reported literature on topological insulators [9 - 11]. The inset of Fig.3 represents the resistivity versus temperature plots under applied magnetic fields [$\rho$(T)H] of 0, 5 and 10 Tesla magnetic fields for the synthesized $Bi_2Se_3$ single crystal. Clearly, the $\rho$-T plots for $Bi_2Se_3$ with and without magnetic field are of metallic nature. The resistivity at 200 K under the absence of magnetic field (0 Tesla) is found to be around 0.52 m$\Omega$-cm and the same increases to 0.62 m$\Omega$-cm and 0.756 m$\Omega$-cm respectively under applied magnetic field of 5 and 10 Tesla. All the $\rho$(T)H plots show positive temperature coefficients, representing that the synthesized single crystal exhibits a metallic behaviour.

Figure 4 represents the Kohler's plot for the synthesized $Bi_2Se_3$ single crystal. According to Kohler's rule the ratio of $\Delta\rho/\rho_o$ and $B/\rho_o$ at different temperatures should consolidate on to a single line in case of single type of charge carriers under applied magnetic fields [12]. Here $\rho_o$ is the residual resistivity at 0 K, as being obtained from the extrapolated $\rho$(T) plot. Interestingly, the Kohler's plot of the magneto - resistance of $Bi_2Se_3$ consolidates onto a single line. Consequently, we can say that the synthesized $Bi_2Se_3$ single crystal does obey Kohler's rule in contrast to previously reported $Bi_2Te_3$ [8].

Figure 5 depicts the Shubnikov-de Haas oscillations (SdH) of $Bi_2Se_3$ single crystal at 2 K under different applied magnetic fields ranging from 4 Tesla to 14 Tesla. Here, the Sdh oscillations plotted as $\Delta\rho/\rho_o$ versus magnetic field at 2 K are calculated by subtracting the experimentally obtained MR (%) versus magnetic field from the linear fitted MR (%). One can observe the Sdh oscillations with increase in magnetic field from say above 8 Tesla to about 14Tesla. As mentioned before, more studies are underway to quantify the SdH oscillations.

Summarily, in current letter we reported the growth, structure and brief magneto transport characterization of self flux grown $Bi_2Se_3$ single crystals. The as grown crystals are large (few cm) in size having metallic conductivity down to 2 K and high non saturating +ve magneto resistance to the tune of above 240% at 2 K in applied field of 14 Tesla. The high MR (250%, 2 K, 14 Tesla) also exhibited the quantum oscillations.

Authors from CSIR-NPL acknowledge the encouragement and support of their Director Prof. D. K. Aswal. Geet Awana thanks Prof. Sanjay Jain, Head Department of Physics and Astrophysics Delhi University for his support.

## Figure Captions

**Figure 1(a):** X-ray diffraction pattern for $Bi_2Se_3$ single crystal, **(b)** Rietveld fitted room temperature X-ray diffraction pattern for powder $Bi_2Se_3$ crystal

**Figure 2:** Scanning electron microscopy image for $Bi_2Se_3$ single crystal, both insets show the elemental analysis.

**Figure 3:** MR (%) as a function of magnetic field for $Bi_2Se_3$ at different temperatures, the inset shows the temperature dependent electrical resistivity for $Bi_2Te_3$ single crystal under different applied magnetic field

**Figure 4:** Kohler plot for $Bi_2Te_3$ in a field range from 0 to 14 Tesla at several temperatures

**Figure 5:** Possible Shubnikov-de Haas oscillations of $Bi_2Se_3$ single crystal at 2 K

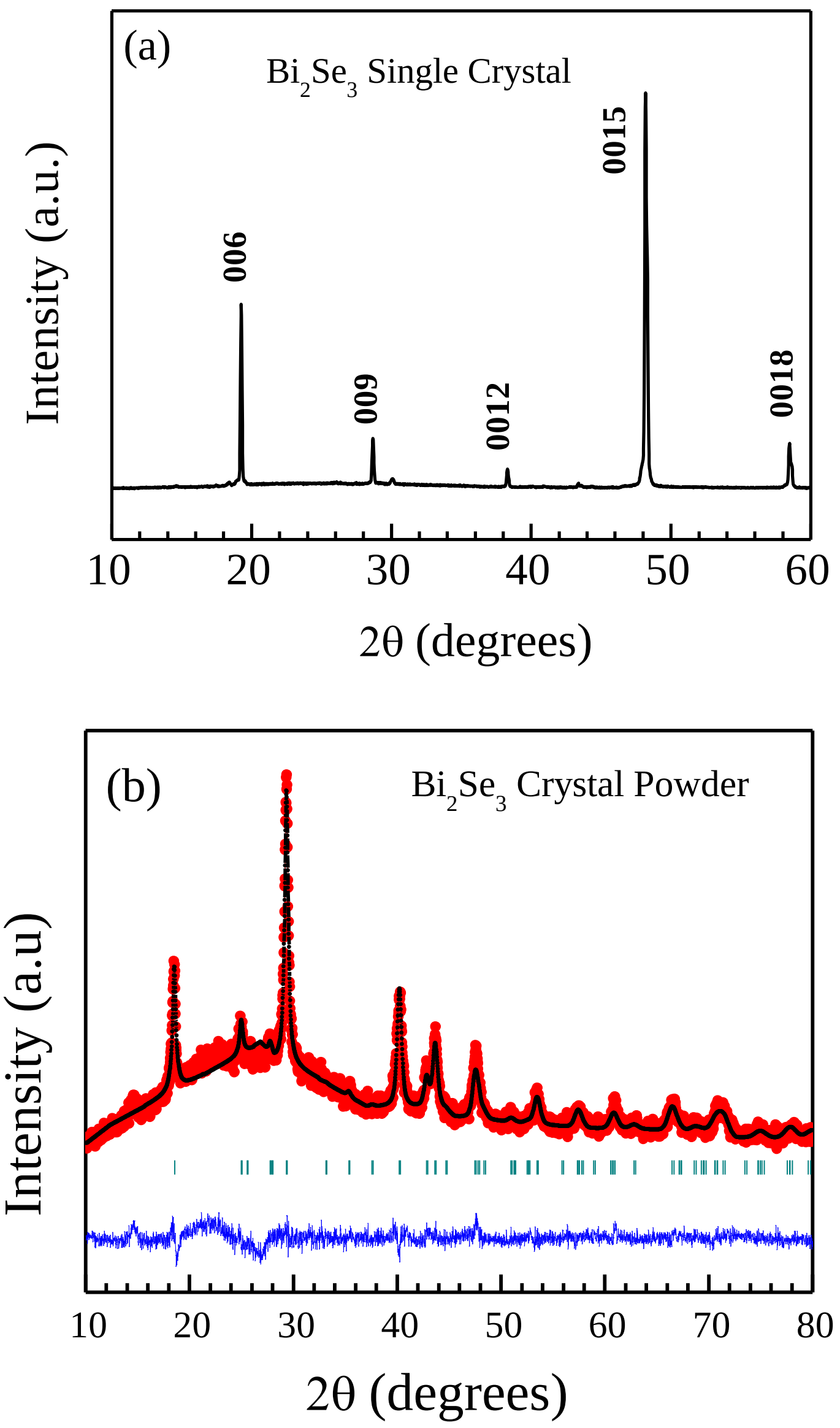

**Fig. 1**

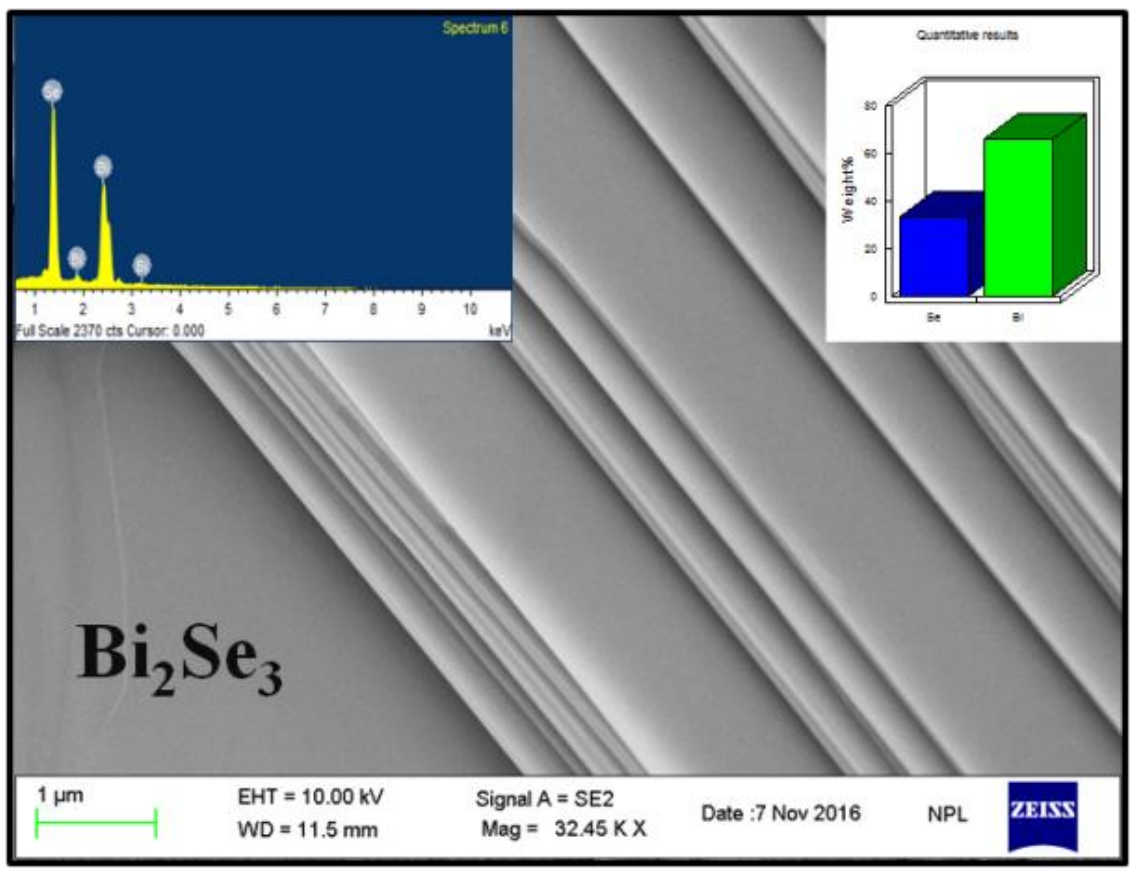

$Bi_2Se_3$
1 µm
EHT = 10.00 kV
WD = 11.5 mm
Signal A = SE2
Mag = 32.45 K X
Date :7 Nov 2016
NPL
ZEISS

**Fig. 2**

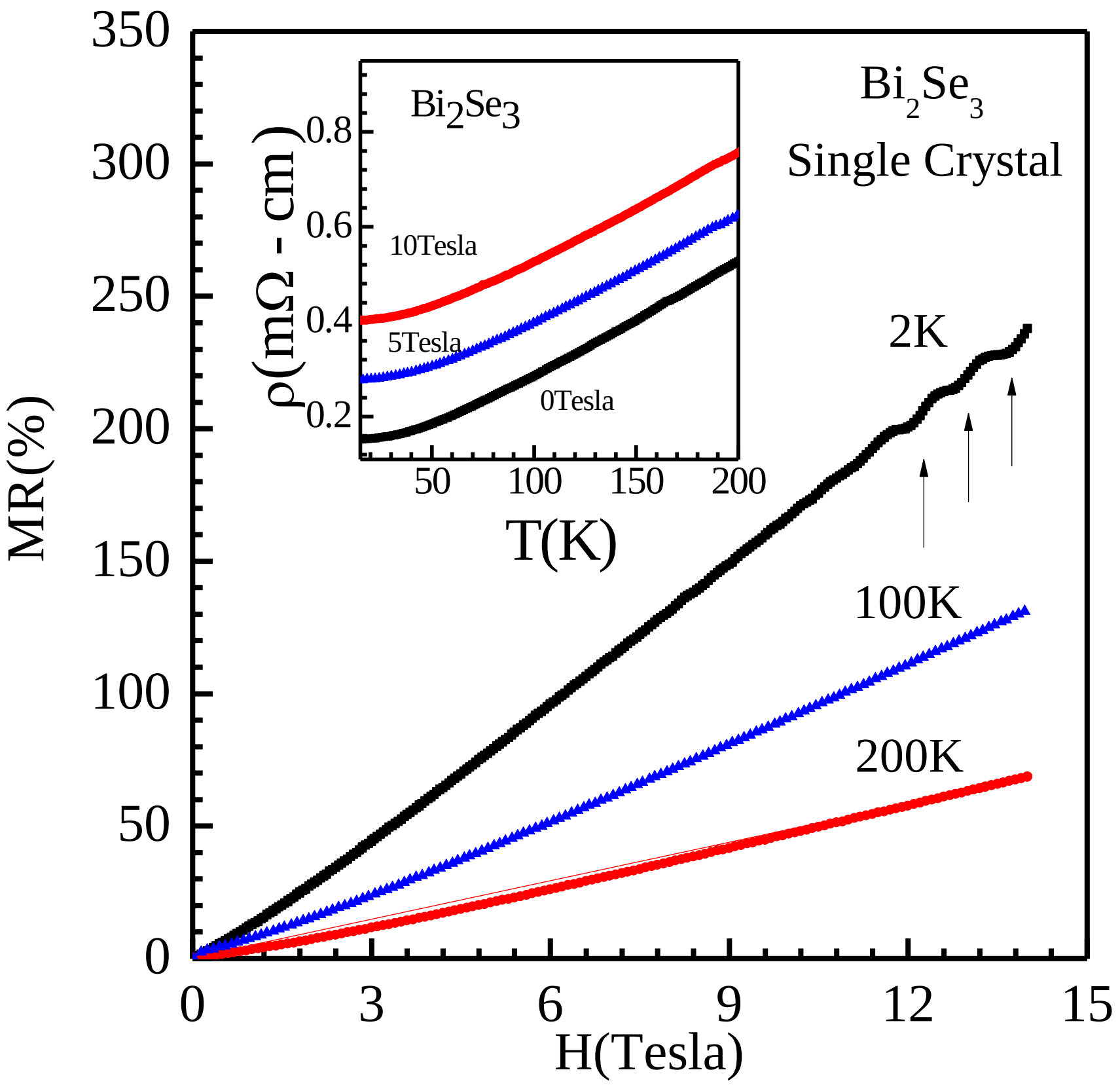

**Fig. 3**

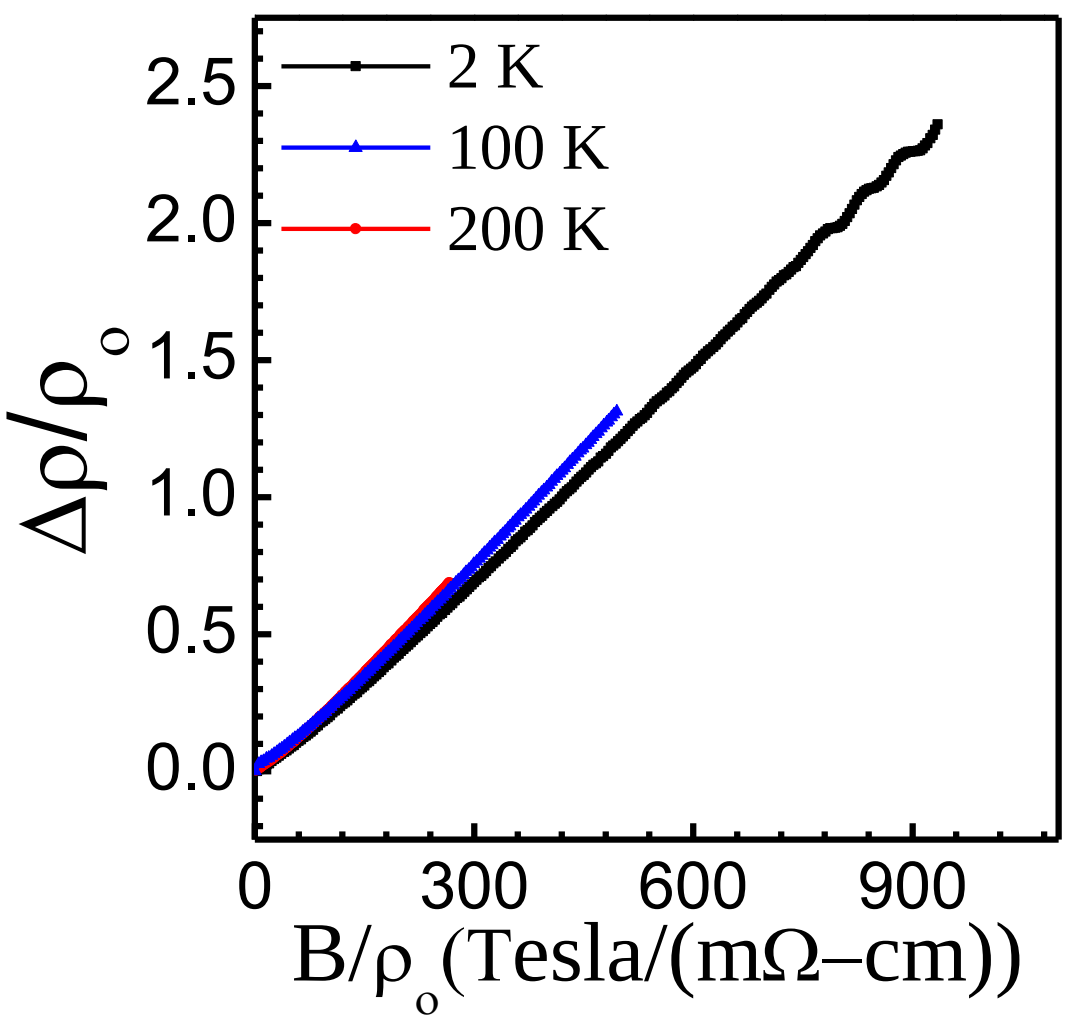

**Fig. 4**

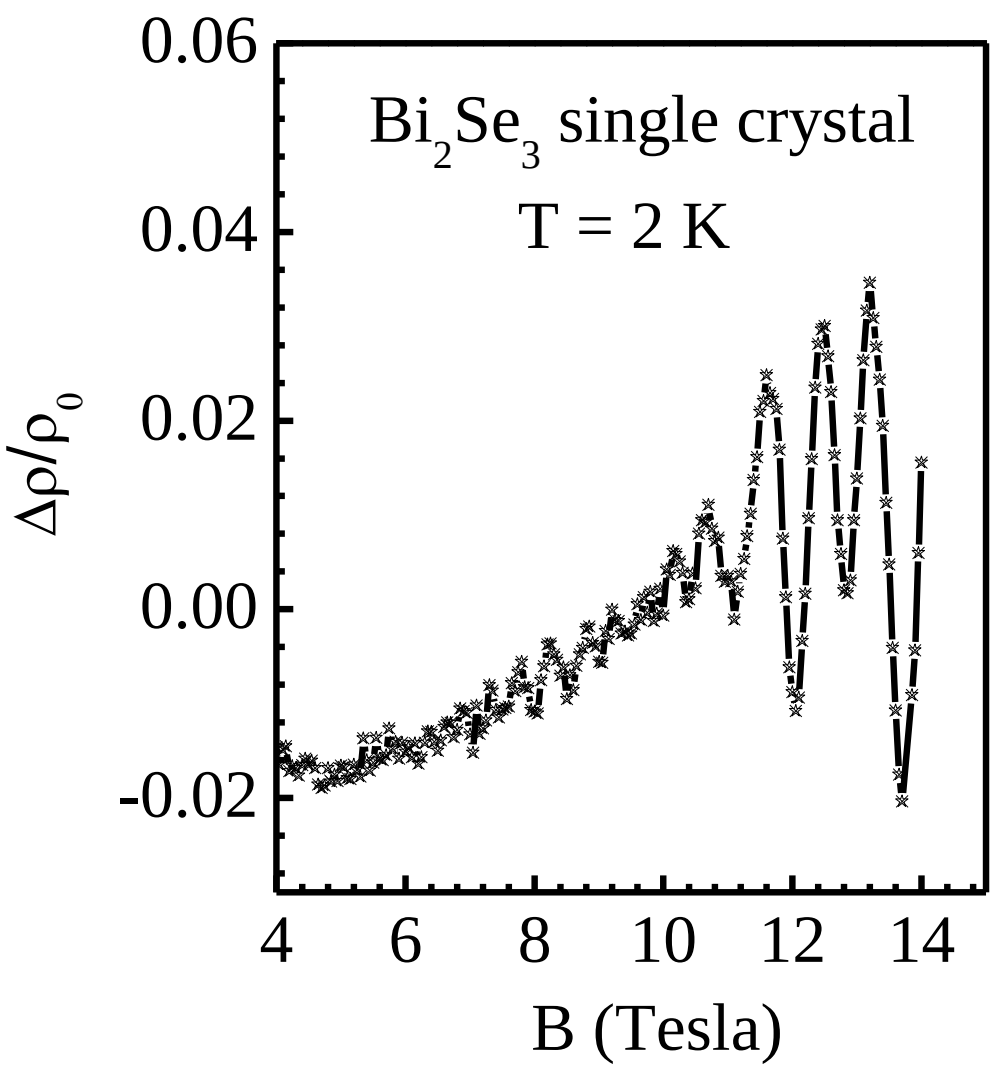

**Fig. 5**